\documentclass[prl,aps,twocolumn,floatfix,amsmath,amssymb,superscriptaddress,tightenlines]{revtex4}
\usepackage{graphicx}
\usepackage{amsfonts}
\usepackage{bm}
\begin{document}

\date{\today}
\title{Valence Bond and von Neumann Entanglement Entropy in Heisenberg Ladders}
\author{Ann B. Kallin}
\affiliation{Department of Physics and Astronomy, University of Waterloo, Ontario, N2L 3G1, Canada} 

\author{Iv\'an Gonz\'alez}
\affiliation{Centro de Supercomputaci\'on de Galicia, Avda.~de~Vigo~s/n, E-15705 Santiago de Compostela, Spain}

\author{Matthew B. Hastings}
\affiliation{Microsoft Research, Station Q, CNSI Building, University of California, Santa Barbara, CA, 93106}

\author{Roger G. Melko}
\affiliation{Department of Physics and Astronomy, University of Waterloo, Ontario, N2L 3G1, Canada} 

\begin{abstract} We present a direct comparison of the recently-proposed
valence bond entanglement entropy and the von Neumann entanglement entropy on
spin 1/2 Heisenberg systems using quantum Monte Carlo and density-matrix
renormalization group simulations.  For one-dimensional chains we
show that the valence bond entropy can be either less or greater than the
von Neumann entropy, hence it cannot provide a bound on the latter.  
On ladder geometries, simulations with up to seven legs are sufficient to indicate that
the von Neumann entropy in two dimensions obeys an area law,
even though the valence bond entanglement entropy 
has a multiplicative logarithmic correction.
\end{abstract}
\maketitle

{\it Introduction.}-- Entanglement has arisen 
as a new paradigm for the study of correlations in condensed matter systems.  
Measurements
of entanglement between subregions, chiefly using entropic
quantities, have an advantage over traditional two-point correlation
functions in that they encode the total amount of information shared
between the subregions without the possibility of missing ``hidden''
correlations \cite{wolf},
such as may occur in some exotic quantum groundstates.   For example spin liquid states, 
where two-point correlation functions decay at large lengthscales, can exhibit topological order that is quantified by a ``topological entanglement
entropy''  \cite{ KP}.
This and other entropic measures are typically discussed in the context of
the von Neumann entanglement entropy ($S^{\rm vN}$), which for a system partitioned into
two regions A and B, quantifies the entanglement between A and B as
\begin{equation} 
S^{\rm vN}_A = - {\rm Tr} [ \rho_A \ln \rho_A ]. \label{vNEE} 
\end{equation}
Here, the reduced density matrix $\rho_A = {\rm Tr}_B | \psi \rangle
\langle \psi |$ is obtained by tracing out the degrees of freedom
of B.

The properties of $S^{ \rm vN}$ are well-studied in 
quantum information theory.
In interacting one-dimensional (1D) quantum systems, exact analytical results are known from conformal
field theories (CFT); they show that, away from special critical points,
$S^{ \rm vN}_A$ scales as the size of the boundary between A and B.
This so-called {\it area law}~\cite{Shredder} is also believed to hold in many
groundstates of two dimensional (2D) interacting quantum systems,
although exact results are scarce~\cite{ALreview}. This has 
consequences in the
rapidly-advancing field of computational quantum many-body theory, where
it is known for example that groundstates of 1D Hamiltonians satisfying an area law
can be accurately represented by matrix product states~\cite{MPS_DMRG}.
Tensor-network states and MERA give two promising new classes of numerical
algorithms~\cite{PEPS1} 
that may allow simulations of 2D quantum systems not amenable to quantum Monte Carlo (QMC) due to
the notorious sign problem.  However, 
these simulation frameworks
are constructed to obey an area law; in order to be represented faithfully
by them, a given 2D quantum groundstate must have 
entanglement
properties also obeying the area law~\cite{ALreview}.

Unfortunately, entanglement is difficult to 
measure in 2D, due to the fact that QMC does not have
direct access to the groundstate wavefunction $| \psi \rangle$ required
in Eq.~\eqref{vNEE}.  In response to this, several authors~\cite{Alet,
Chh} have introduced the concept of {\it valence bond}
entanglement entropy ($S^{\rm VB}$), defined for a  spin system as
\begin{equation} 
S^{\rm VB}_A = \ln(2) \cdot \mathcal{N}_A, \label{VBEE}
\end{equation} 
where $ \mathcal{N}_A$ is the number of singlets
${( |\uparrow \downarrow \rangle - | \downarrow \uparrow
\rangle)/\sqrt{2}}$ crossing the boundary between regions A and B.  Unlike
$S^{ \rm vN}$, $S^{\rm VB}$ can be accessed easily in the valence-bond basis
projector QMC method recently proposed by Sandvik \cite{Sandvik}.  As
demonstrated in Refs.~\cite{Alet,Chh}, $S^{\rm VB}$ has many properties in
common with $S^{ \rm vN}$, in particular the relationship $S^{\rm VB}_A = S^{\rm VB}_B$, and the
fact that $S^{\rm VB}_A=0$ for regions ``un-entangled'' by valence bonds.
A comparison of the scaling of $S^{\rm VB}$ for (critical) 1D spin
1/2 Heisenberg chains shows good agreement with analytical results known
from CFT, however in the
 2D isotropic Heisenberg model it
displays a {\it multiplicative} logarithmic correction to the area law~\cite{Alet,Chh}.  If
true also for $S^{ \rm vN}$, this would have negative consequences for the simulation of the 
N\'eel groundstate using tensor-network simulations.
 
In this paper, we compare $S^{\rm VB}$ calculated by valence-bond QMC to 
$S^{ \rm vN}$ accessible through density-matrix renormalization group
(DMRG) simulations of the Heisenberg model on $N$-leg ladders.    For $N=1$, the CFT central charge calculated  by scaling
$S^{ \rm vN}$ shows excellent agreement to $c=1$, whereas $S^{\rm VB}$ converges
to $c<1$.
For $N>1$, $S^{\rm VB}$ is systematically greater than $S^{ \rm vN}$,
a trend which grows rapidly with $N$. An
examination of entanglement defined by bipartitioning multi-leg ladders
shows a logarithmic correction for the valence-bond entanglement entropy,
$S^{\rm VB} \propto N \ln
N$ (in agreement with Refs.~\cite{Alet,Chh}), however for data up to
$N=7$, the von Neumann entanglement entropy  convincingly shows a scaling of
$S^{\rm vN} \propto N$, thus obeying the area law.

{\it Model and Methods.}-- We consider the spin 1/2 Heisenberg
Hamiltonian, given by  $H =  \sum_{\langle i j \rangle} {\mathbf S}_i
\cdot {\mathbf S}_j \label{ham}$, where the sum is over nearest-neighbor
lattice sites.  Geometries studied are 1D chains, and multi-leg ladders
with length $L$ and number of legs $N$.  
We employ two complementary numerical techniques, the valence-bond basis
QMC and DMRG, both of which give {\it unbiased} approximations to the
ground state of the Hamiltonian at zero temperature.  $S^{ \rm vN}_A$ is
naturally accessible through the DMRG algorithm~\cite{White92}.  At each
step of the algorithm, the wavefunction of the system is approximated by
keeping only the states with largest coefficients in the Schmidt
decomposition for a given bipartition into regions A and
B~$\equiv\complement$A. To find the basis entering the Schmidt
decomposition for region A, the reduced density matrix $\rho_A$ is
calculated and diagonalized, thus allowing easy calculation of
Eq.~\eqref{vNEE}. The truncation of the basis implies that only a lower
bound for $S^{\rm vN}_A$ is calculated, so care must be taken to ensure
that enough of the eigenvalue spectrum is included to converge $S^{ \rm
vN}_A$ to sufficient accuracy; typically the number of states required is
larger than that needed to converge the energy alone.
We have kept up to 1800 states per block for the largest ladders, with
truncation errors under 10$^{-7}$ in all cases.

$S^{\rm VB}_A$ 
can be calculated using the valence-bond basis
QMC proposed by Sandvik \cite{Sandvik}.  The 
algorithm we use is the simple single-projector method, with lattice
geometries constructed to match those used by DMRG.
The
ground state of the system is projected out by repeated application of a
list of nearest-neighbor bond operators,
a number
of which are changed each step.  The change is accepted
with a probability of $2^{n_{b}}/2^{n_{a}} $ where $n_{a}$ ($n_{b}$)
is the number of off-diagonal operators in the current (last accepted) step.
Measured quantities such as energy or $S^{\rm VB}_A$ are then calculated 
by a weighted average over all the valence bond
states obtained by this procedure.

\begin{figure} {
\includegraphics[width=3.3in]{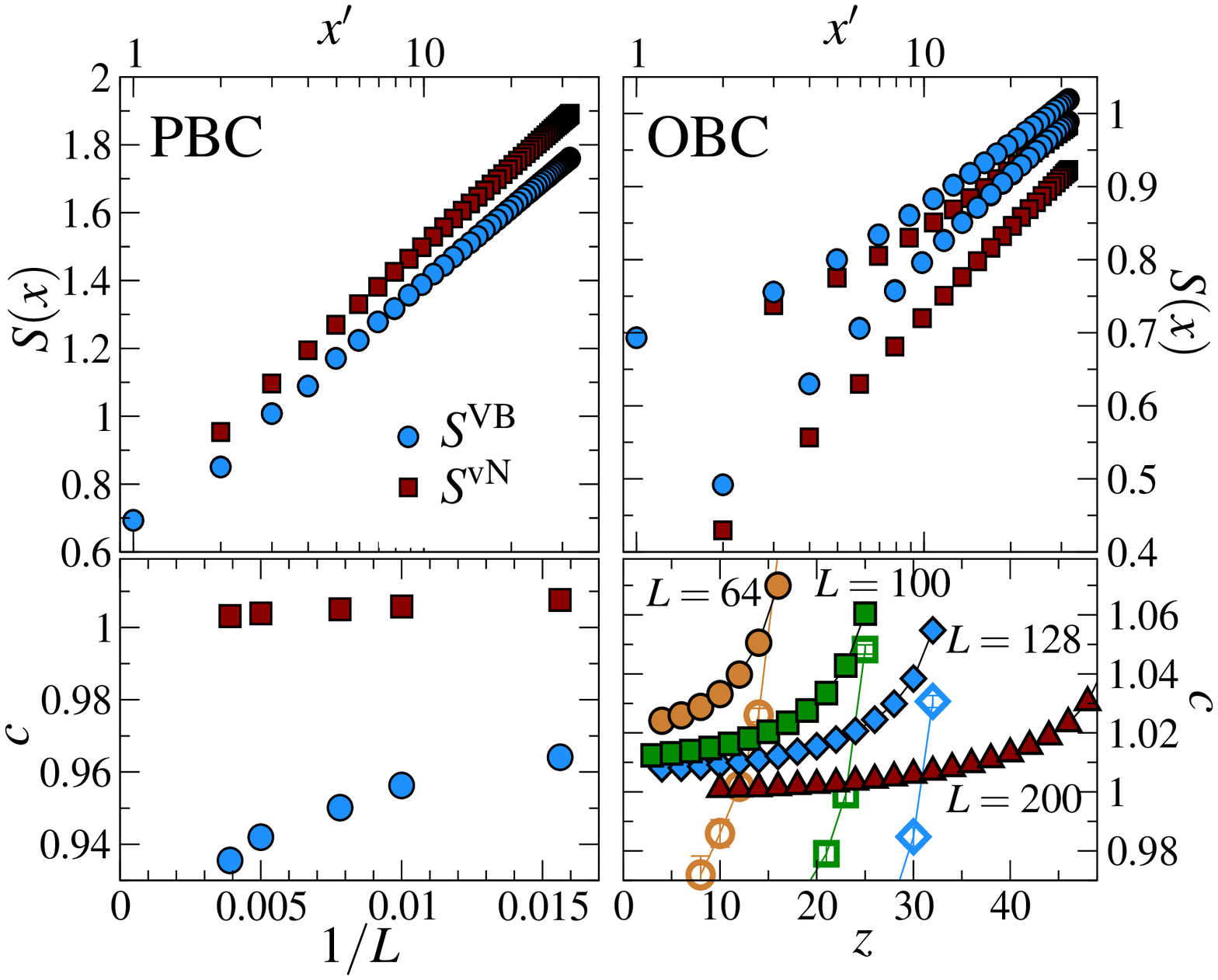} \caption{(color online) 
Entanglement entropies for a 1D Heisenberg chain with PBC and OBC. Upper
panels show the entropies as a function of the conformal distance $x'  = (L/\pi)\sin (\pi x/L)$ for
100-site chains.  Lower plots show the central
charge $c$, obtained by fitting the numerical data to the CFT result, for
several $L$.  For PBC, $c$ is calculated with the two smallest $x'$ points
removed.  For OBC, the fits depend on the number of sites included, $z$,
which we systematically decrease by removing $x'$ data points from the
{\it outside} ends of the chain.  $c$ is shown for $S^{\rm vN}$
(closed symbols) and $S^{\rm VB}$ (open symbols) for system sizes $L=64$
(circles), $L=100$ (squares), $L=128$ (diamonds), and $L=200$ (triangles)
\label{1D}}} \end{figure}

{\it One-dimensional chain.}-- We consider first the case of Heisenberg
chains ($N=1$) of length $L$, simulating both open (OBC) and periodic
(PBC) boundary conditions.  The DMRG algorithm requires the regions A and
B to be topologically connected, so in 1D the bipartition is defined by a
site index $x$ with sites within the interval $[1,x]$ ($[x+1,L]$)
belonging to region $A$ ($B$) [thus we can label $S_A$ by its site index,
$S(x)$].  We stress that the QMC and DMRG results are on the same geometry
and Hamiltonian, and reproduce the same ground state energies; Figure 1
(and subsequent figures) should be considered as exact comparisons between
$S^{\rm VB}$ and $S^{ \rm vN}$.

The 1D Heisenberg model is known to be critical and thus can be
mapped to a 2D classical Hamiltonian at its critical point, which
can be described by CFT in the limit $L\to\infty$.  To address
finite-length chains one can use the conformal mappings $x\to x'=(L/\pi)
\sin(\pi x / L)$ for PBC, and 
$x\to 2x'$ for OBC. 
Calculations within CFT~\cite{Cardy} obey $S^{\rm vN}(x)= (c/3)
\ln(x') + S_1$ for PBC, and $S^{\rm vN}(x)= (c/6) \ln(2x') +
\ln(g)+S_1/2$ for OBC, where $c$ is the central charge of the CFT,
$S_1$ is a model-dependent constant, and $g$ is Ludwig and Affleck's
universal boundary term~\cite{AffleckAndLudwig}.

Figure~\ref{1D} illustrates simulation results in both cases, the left
(right) panels corresponding to PBC (OBC). 
For PBC both
$S^{\rm VB}$ and $S^{ \rm vN}$ appear to fit well to the CFT result, although
$S^{ \rm vN} > S^{\rm VB}$. The regression fit for $S^{ \rm vN}$ shows very good
convergence with the central charge predicted by CFT, $c=1$, while
the fit for $S^{\rm VB}$ yields a lower value of $c$ than predicted.
For OBC both $S^{ \rm vN}$ and $S^{\rm VB}$ split into two branches, the upper (lower)
corresponding to an odd (even) number of lattice sites in A.  This
reflects a well-known ``dimerization'' effect induced by OBC~\cite{Ian1}.
Notice that contrary to the PBC case, now $S^{ \rm vN} < S^{\rm VB}$. 
A regression fit of the lower branch to the form $(c/6) \ln
({2x'})$ (inset) shows excellent convergence of $S^{ \rm vN}$ to the central
charge predicted by CFT, $c=1$, once finite-size effects and the proximity
of the data to the open boundaries are taken into account.  In contrast,
 $S^{\rm VB}$ deviates significantly from the CFT result, giving 
$c>1$ when all or most data is included in the fit, changing to $c<1$
as data closest to the open boundary is systematically excluded~\cite{XXX}, e.g. for $z=L/2$, $c_{L\to\infty} \approx 0.85$.

\begin{figure} { \includegraphics[width=3.3in]{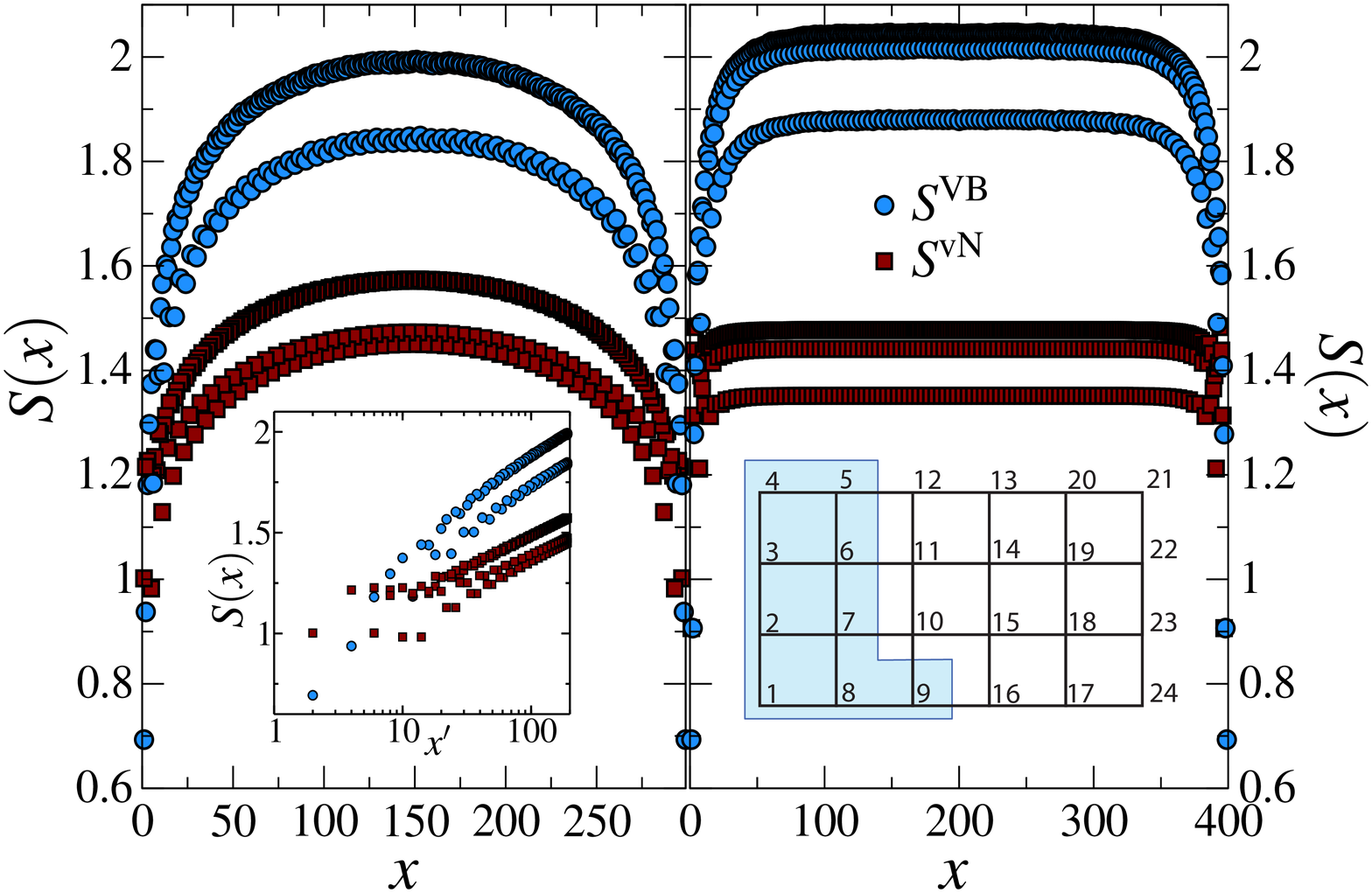}
\caption{(color online) Entanglement entropies for 3-leg (left)
and 4-leg (right) ladder systems with OBC and 100 sites per leg.  For
odd-leg ladders, $S(x)\propto\ln(x')$.  The left
inset shows $S(x)$ as a function of the conformal distance, $x'$, on a log
scale. For even-leg ladders, $S(x\gtrsim\xi)= {\rm const}.$
The right inset shows the site indexing used for multi-leg ladders where the
bipartition A is shaded and labeled by $x=9$.  \label{ladder} }} \end{figure}

{\it Multi-leg ladders.}-- Moving away from the 1D chain, one can add
``legs'' to the lattice in a systematic way. In this case the sum over
nearest neighbors is extended to neighbors along rungs as well as along
legs.  As noted before, DMRG imposes constraints to the subregion
geometry. In multi-leg ladders we choose to sweep in a 1D path that visits
first bonds sitting in rungs rather than bonds sitting in legs (see
Fig.~\ref{ladder}).  DMRG computational demands increase dramatically with
the number of legs, so in this paper we restrict ourselves to ladders with
OBC up to $N=7$ legs. QMC lacks this limitation and one can go up to
$N=20$ with minimal CPU effort.

Figure~\ref{ladder} shows $S^{ \rm vN}$ and $S^{\rm VB}$ calculated for
the 3-leg and 4-leg ladder. As for the OBC 1D chain,~$S^{\rm VB}>S^{ \rm vN}$.  
Entropy shows different behavior depending on $N$ being even or odd.
Even-leg ladders have a spin-gap~\cite{White1994}, and thus only
sites within distances from the boundary between A and B {\it smaller}
than the correlation length $\xi$ contribute to the entanglement, yielding
$S(x\gtrsim \xi)= {\rm const}.$ In contrast, odd-leg ladders are gapless,
and thus all sites 
contribute to the
entanglement, yielding $S(x)\propto\ln(x')$, which follows the CFT result
in analogy to the 1D case. As can be seen, $S(x)$ splits into branches,
with a (quasi-) periodic structure superimposed over the main dependence
on $x$, the period being $N$ ($2N$) for even- (odd-) leg ladders. This
reflects the periodicity of the underlying 1D path through which the
algorithm sweeps, and the fact that valence bonds within the same rung are
energetically favored~\cite{White1994}. The doubled period for odd-leg
ladders is due to the same dimerization effect as in Ref.~\cite{Ian1}.

{\it Area law in multi-leg ladders.}--  
We can use these results to address the question of the adherence of the 2D N\'eel state
to an area law.  To do so, we
define the lattice
geometry such that region A is rectangular, cutting a multi-leg ladder
cleanly across all legs, so that the ``area'' separating region A and B is
equivalent to the number of legs in the ladder $N$.  We choose the region A
to contain $2N^2$ sites, to have an aspect ratio of order unity.  In
contrast, the entanglement entropy of a long
narrow region would be dominated by the behavior
of the gapless mode for odd-leg ladders.
The 2:1 aspect ratio makes the region length even for all $N$, reducing even-odd oscillations.

\begin{figure} { \includegraphics[width=3in]{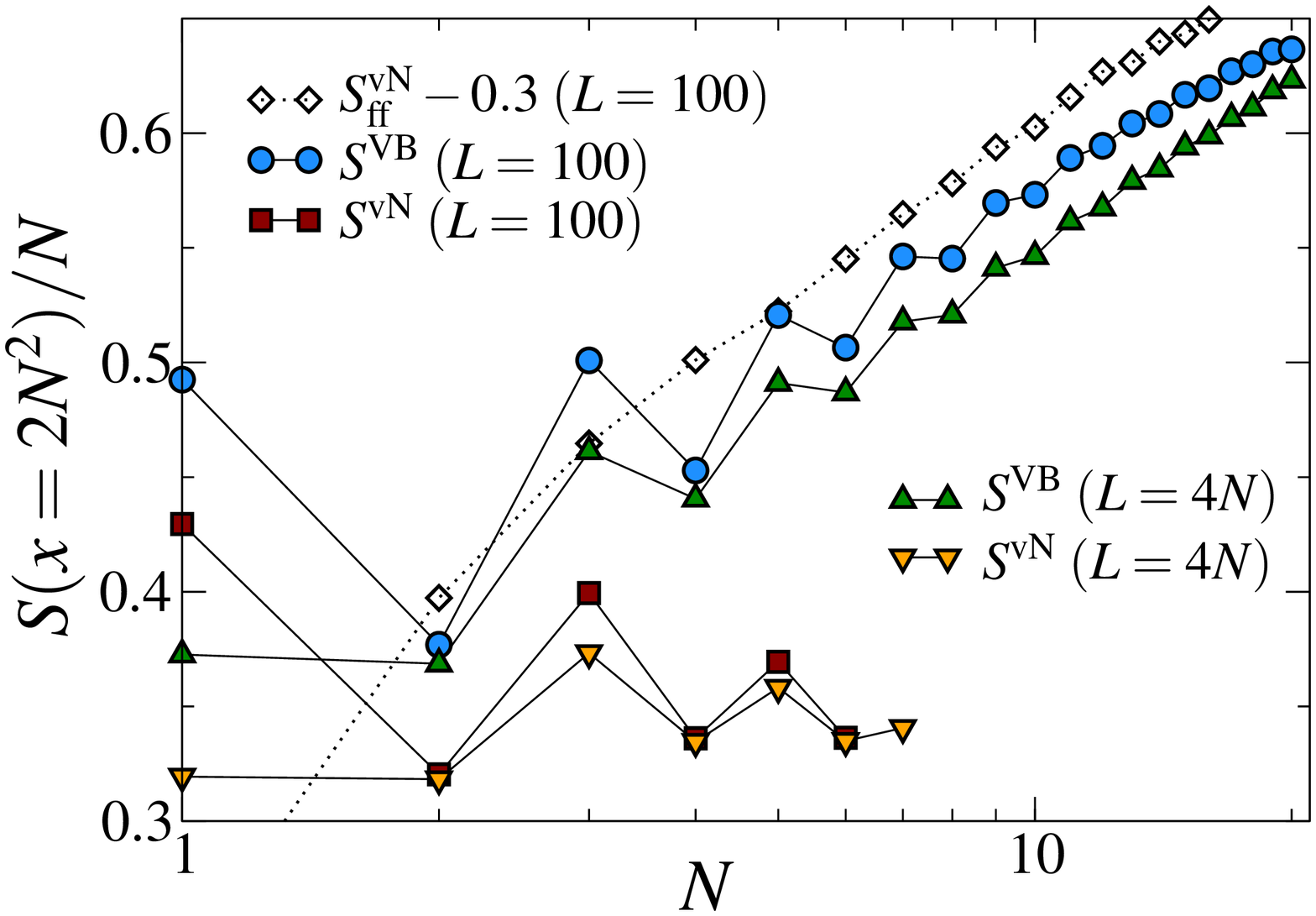} \caption{(color
online) Entanglement entropies divided by $N$,  for $N$-leg Heisenberg
(filled symbols) and free-fermion (open diamonds) ladders, taken such that
the region A includes $2N^2$ sites.  
For the Heisenberg model and large $N$, $S^{\rm VB}\propto N \ln N$,
whereas $S^{\rm vN}\propto N$.  
Data for free fermions, $S^{\rm vN}_{\rm ff}\propto N \ln N$,  are shown for comparison.
We show data for ladders with length (sites per leg) $L =100$ and
$L=4N$.  \label{zigzag}}} \end{figure}

Figure~\ref{zigzag} illustrates the simulation results for $N$-leg ladders.
Plotting $S(x)/N$ versus $N$ on a log scale, we obtain a
multiplicative logarithmic correction to the $S^{\rm VB}$ area law, in agreement
with results from Refs.~\cite{Alet,Chh}.  However, the linear slope is {\it not} present in
the plot of the $S^{ \rm vN}$ data from the DMRG, which convincingly approaches a
constant for large $N$ \cite{Snote}.  Clearly, for $S^{ \rm vN}$ the area law is
indeed obeyed in the N\'eel groundstate, leading one to conclude that the
multiplicative logarithmic correction occurs in $S^{\rm VB}$ only.  One can
compare $S^{\rm VB}$ to data obtained for free fermions, which has a well-known \cite{2dfermion} logarithmic correction to the area law for $S^{ \rm vN}$ (Fig.~\ref{zigzag}).
In the next section, we explain $S^{\rm VB}$
in the context of the bond length distribution in the QMC.
We note also that, contrary to the suggestion in Ref.~\cite{Alet}, a gapless Goldstone mode will {\it not} give a logarithmic
divergence to $S^{ \rm vN}$ since a gapless bosonic mode in 2D
obeys an area law \cite{2dboson}.  We have also done a spin-wave calculation of
$S^{\rm vN}$ for this system and found an area law, albeit with
$S(x)/N\approx 0.2$, slightly lower than suggested for spin~$1/2$ in Fig.~\ref{zigzag}.

{\it Bond Length Distribution.}-- Sandvik defined the bond length
distribution $P(x,y)$ as the probability of a bond going from site $x$ to
site $y$, and found that $P(x,y)\sim |x-y|^{-p}$ with $p\approx 3$ in the
N\'eel state~\cite{Sandvik}.  This value of $p$ gives the logarithmic
divergence in $S^{\rm VB}$, as can be found by directly calculating
\begin{equation} 
S^{\rm VB}_{A}=\sum_{x\in A,y\in B} P(x,y)\ln (2).
\end{equation}

We can understand the value of $p$ from a scaling argument similar to that
in Ref.~\cite{network}.
Consider the number of bonds of length $l$ exiting a region of linear size $l$.  For $p>4$, this
scales to zero and such a state should have no long-range order.
For $p=4$, this number is $l$-independent, corresponding to a critical
state ($p=2$ is the critical power in 1D, matching the
observed logarithmic behavior of $S^{\rm VB}$).
For $p<2$, the bond length distribution is unnormalized, and all bonds are long, so that
there is no short-range order.  $p=3$ represents a state with both long-
and short-range order.  There is a heuristic argument that $p=3$ corresponds to a state
with Goldstone modes: the correlation function of two spins is the
probability that they lie on the same loop.  This can be
estimated by
the Green's function of a L{\'e}vy flight with power law distributed steps with power $3$,
reproducing the power law decay of correlations.

{\it Discussion.}-- In this paper, we have compared scaling properties of
the valence bond entanglement entropy ($S^{\rm VB}$) \cite{Alet,Chh} to the von Neumann entanglement entropy ($S^{ \rm vN}$) in the
spin 1/2 Heisenberg model on multi-leg ladder geometries, using QMC and DMRG simulations.
In 1D, we find
that $S^{\rm VB}$ mimics the behavior of $S^{ \rm vN}$ closely, although 
it is less than $S^{ \rm vN}$ for periodic
chains, and greater than $S^{ \rm vN}$ for open chains. In addition, fits to
1D conformal field theory, which are excellent for $S^{ \rm vN}$ calculated
via DMRG, appear to deviate significantly for $S^{\rm VB}$ in the large
chain-size limit, approaching $c<1$ for both boundary conditions~\cite{XXX}.

The fact  that $S^{\rm VB}$ can be either greater or less than $S^{ \rm
vN}$ can be understood through simple examples. Let $|(ij)(kl)...\rangle$
denote a state in which sites $i,j$ are in a singlet, sites $k,l$ are in a
singlet, and so on.  Consider an 8-site chain, with sites 1 to 4 in region
A. Then, the state $|(12)(34)(56)(78)\rangle+|(14)(32)(58)(76)\rangle$ has
vanishing $S^{\rm VB}_A$ since no bonds connect A to B, but
non-vanishing $S^{ \rm vN}_A \approx 0.325$.  On the other hand, consider a 4-site chain,
with sites $1$ and $3$ in region A. Then, the state
$|(12)(34)\rangle+|(14)(32)\rangle$ has a maximal $S^{\rm VB}_A$, equal to
$2\ln(2) \approx 1.386$, while $S^{ \rm vN}_A = \ln(3) \approx 1.099$ is smaller.  This second state
has the maximum possible N\'eel order parameter: it is the
equal amplitude superposition of all configurations of bonds connecting the
two sublattices.
Thus, it is unsurprising that states with N\'eel order show $S^{\rm VB} > S^{ \rm vN}$,
a fact which we have demonstrated numerically on multi-leg ladder systems with open boundaries.

Defining the boundary between the two entangled regions as being bipartitioned by a cut across all legs
on a ladder, we have shown using DMRG that $S^{ \rm vN}$ obeys the area law in the many-leg limit.   
Since DMRG can also accurately measure logarithmic size-dependencies of $S^{\rm vN}$ (in 1D critical systems), this suggests that simulation procedures similar to those here might enable the measurement of area-law corrections in $S^{\rm vN}$ as indicators of exotic phases in other models, such as those with
a spinon Fermi surface~\cite{BoseMetal}.

The valence bond entanglement entropy harbors a multiplicative logarithmic
correction for the N\'eel groundstate, which we have shown
is caused by the valence-bond length distribution, and is not present in
$S^{\rm vN}$.  It is clear that $S^{\rm VB}$ is a reasonable
measurement of entanglement, readily accessible to numerical simulations
in 2D and higher, and capable of reproducing the area law in some gapped
groundstates \cite{Alet,Chh}.  However, the inability of  $S^{\rm VB}$ to
provide a bound on $S^{\rm vN}$ (unlike other measures such as Renyi
entropies), along with its discrepancies from $S^{\rm vN}$ in 1D critical
systems and the 2D N\'eel state, must be taken into account in proposals
to use $S^{\rm VB}$ for
future tasks such as characterizing topological phases or studying
universality at quantum phase transitions.

{\it Acknowledgments.}-- The authors thank I.~Affleck, A.~J.~Berlinsky,
N.~Bonesteel, A.~Del~Maestro, A.~Feiguin, 
M.~Fisher, L.~Hormozi, and E.~S\o rensen 
for useful discussions.  This work was made possible by the
computing facilities of SHARCNET and CESGA.  Support was provided by NSERC
of Canada (A.B.K. and R.G.M.) and the NSF under Grant No. NSF PHY05-51164
(I.G.).

\bibliography{VB_biblio}

\begin{thebibliography}{99}
\expandafter\ifx\csname natexlab\endcsname\relax\def\natexlab#1{#1}\fi
\expandafter\ifx\csname bibnamefont\endcsname\relax
  \def\bibnamefont#1{#1}\fi
\expandafter\ifx\csname bibfnamefont\endcsname\relax
  \def\bibfnamefont#1{#1}\fi
\expandafter\ifx\csname citenamefont\endcsname\relax
  \def\citenamefont#1{#1}\fi
\expandafter\ifx\csname url\endcsname\relax
  \def\url#1{\texttt{#1}}\fi
\expandafter\ifx\csname urlprefix\endcsname\relax\def\urlprefix{URL }\fi
\providecommand{\bibinfo}[2]{#2}
\providecommand{\eprint}[2][]{\url{#2}}

\bibitem[{\citenamefont{Wolf et~al.}(2008)\citenamefont{Wolf, Verstraete,
  Hastings, and Cirac}}]{wolf}
\bibinfo{author}{\bibfnamefont{M.~M.} \bibnamefont{Wolf}},
  \bibinfo{author}{\bibfnamefont{F.}~\bibnamefont{Verstraete}},
  \bibinfo{author}{\bibfnamefont{M.~B.} \bibnamefont{Hastings}},
  \bibnamefont{and} \bibinfo{author}{\bibfnamefont{J.~I.} \bibnamefont{Cirac}},
  \bibinfo{journal}{Phys. Rev. Lett.} \textbf{\bibinfo{volume}{100}},
  \bibinfo{pages}{070502} (\bibinfo{year}{2008}).

\bibitem[{\citenamefont{Kitaev and Preskill}(2006)}]{KP}
\bibinfo{author}{\bibfnamefont{A.}~\bibnamefont{Kitaev}} \bibnamefont{and}
  \bibinfo{author}{\bibfnamefont{J.}~\bibnamefont{Preskill}},
  \bibinfo{journal}{Phys. Rev. Lett.} 
  \textbf{\bibinfo{volume}{96}},
  \bibinfo{eid}{110404} (\bibinfo{year}{2006});
\bibinfo{author}{\bibfnamefont{M.}~\bibnamefont{Levin}} \bibnamefont{and}
  \bibinfo{author}{\bibfnamefont{X.-G.} \bibnamefont{Wen}},
  \bibinfo{journal}{{\it ibid.}} \textbf{\bibinfo{volume}{96}},
  \bibinfo{eid}{110405} (\bibinfo{year}{2006}).

\bibitem[{\citenamefont{Srednicki}(1993)}]{Shredder}
\bibinfo{author}{\bibfnamefont{M.}~\bibnamefont{Srednicki}},
  \bibinfo{journal}{Phys. Rev. Lett.} \textbf{\bibinfo{volume}{71}},
  \bibinfo{pages}{666} (\bibinfo{year}{1993}).

\bibitem[{\citenamefont{Eisert et~al.}(2008)\citenamefont{Eisert, Cramer, and
  Plenio}}]{ALreview}
\bibinfo{author}{\bibfnamefont{J.}~\bibnamefont{Eisert}},
  \bibinfo{author}{\bibfnamefont{M.}~\bibnamefont{Cramer}}, \bibnamefont{and}
  \bibinfo{author}{\bibfnamefont{M.}~\bibnamefont{Plenio}},
  \eprint{arXiv:0808.3773}.

\bibitem[{\citenamefont{\"Ostlund and Rommer}(1995)}]{MPS_DMRG}
\bibinfo{author}{\bibfnamefont{S.}~\bibnamefont{\"Ostlund}} \bibnamefont{and}
  \bibinfo{author}{\bibfnamefont{S.}~\bibnamefont{Rommer}},
  \bibinfo{journal}{Phys. Rev. Lett.} \textbf{\bibinfo{volume}{75}},
  \bibinfo{pages}{3537} (\bibinfo{year}{1995}).

\bibitem[{\citenamefont{Verstraete and Cirac}(2004)}]{PEPS1}
\bibinfo{author}{\bibfnamefont{F.}~\bibnamefont{Verstraete}} \bibnamefont{and}
  \bibinfo{author}{\bibfnamefont{J.~I.} \bibnamefont{Cirac}},
  \eprint{arXiv:cond-mat/0407066};
\bibinfo{author}{\bibfnamefont{F.}~\bibnamefont{Verstraete}},
  \bibinfo{author}{\bibfnamefont{M.~M.} \bibnamefont{Wolf}},
  \bibinfo{author}{\bibfnamefont{D.}~\bibnamefont{Perez-Garcia}},
  \bibnamefont{and} \bibinfo{author}{\bibfnamefont{J.~I.} \bibnamefont{Cirac}},
  \bibinfo{journal}{Phys. Rev. Lett.} \textbf{\bibinfo{volume}{96}},
  \bibinfo{pages}{220601} (\bibinfo{year}{2006});
\bibinfo{author}{\bibfnamefont{G.}~\bibnamefont{Evenbly}} \bibnamefont{and}
  \bibinfo{author}{\bibfnamefont{G.}~\bibnamefont{Vidal}},
  \bibinfo{journal}{Phys. Rev. B} \textbf{\bibinfo{volume}{79}},
  \bibinfo{pages}{144108} (\bibinfo{year}{2009}).

\bibitem[{\citenamefont{Alet et~al.}(2007)\citenamefont{Alet, Capponi,
  Laflorencie, and Mambrini}}]{Alet}
\bibinfo{author}{\bibfnamefont{F.}~\bibnamefont{Alet}},
  \bibinfo{author}{\bibfnamefont{S.}~\bibnamefont{Capponi}},
  \bibinfo{author}{\bibfnamefont{N.}~\bibnamefont{Laflorencie}},
  \bibnamefont{and} \bibinfo{author}{\bibfnamefont{M.}~\bibnamefont{Mambrini}},
  \bibinfo{journal}{Phys. Rev. Lett.} \textbf{\bibinfo{volume}{99}},
  \bibinfo{pages}{117204} (\bibinfo{year}{2007}).

\bibitem[{\citenamefont{Chhajlany et~al.}(2007)\citenamefont{Chhajlany,
  Tomczak, and W\'{o}jcik}}]{Chh}
\bibinfo{author}{\bibfnamefont{R.~W.} \bibnamefont{Chhajlany}},
  \bibinfo{author}{\bibfnamefont{P.}~\bibnamefont{Tomczak}}, \bibnamefont{and}
  \bibinfo{author}{\bibfnamefont{A.}~\bibnamefont{W\'{o}jcik}},
  \bibinfo{journal}{Phys. Rev. Lett.} \textbf{\bibinfo{volume}{99}},
  \bibinfo{eid}{167204} (\bibinfo{year}{2007}).

\bibitem[{\citenamefont{Sandvik}(2005)}]{Sandvik}
\bibinfo{author}{\bibfnamefont{A.~W.} \bibnamefont{Sandvik}},
  \bibinfo{journal}{Phys. Rev. Lett.} \textbf{\bibinfo{volume}{95}},
  \bibinfo{pages}{207203} (\bibinfo{year}{2005}).

\bibitem[{\citenamefont{White}(1992)}]{White92}
\bibinfo{author}{\bibfnamefont{S.~R.} \bibnamefont{White}},
  \bibinfo{journal}{Phys. Rev. Lett.} \textbf{\bibinfo{volume}{69}},
  \bibinfo{pages}{2863} (\bibinfo{year}{1992});
  \bibinfo{author}{\bibfnamefont{S.~R.} \bibnamefont{White}},
  \bibinfo{journal}{Phys. Rev. B} \textbf{\bibinfo{volume}{48}},
  \bibinfo{pages}{10345} (\bibinfo{year}{1993}).  See also
\bibinfo{author}{\bibfnamefont{U.}~\bibnamefont{Schollw\"ock}},
  \bibinfo{journal}{Rev. Mod. Phys.} \textbf{\bibinfo{volume}{77}},
  \bibinfo{pages}{259} (\bibinfo{year}{2005}).

\bibitem[{\citenamefont{Calabrese and Cardy}(2004)}]{Cardy}
\bibinfo{author}{\bibfnamefont{P.}~\bibnamefont{Calabrese}} \bibnamefont{and}
  \bibinfo{author}{\bibfnamefont{J.}~\bibnamefont{Cardy}}, \bibinfo{journal}{J.
  Stat. Mech.: Theor. Exp.} {\bibinfo{volume}{P06002}}
  (\bibinfo{year}{2004});
  \bibinfo{author}{\bibfnamefont{H.~Q.}~\bibnamefont{Zhou}},
  \bibinfo{author}{\bibfnamefont{T.}~\bibnamefont{Barthel}}, 
  \bibinfo{author}{\bibfnamefont{J.~O.}~\bibnamefont{Fjaerestad}}, \bibnamefont{and}
  \bibinfo{author}{\bibfnamefont{U.}~\bibnamefont{Schollwoeck}},
  \bibinfo{journal}{Phys. Rev. A} \textbf{\bibinfo{volume}{74}},
  \bibinfo{pages}{050305(R)} (\bibinfo{year}{2006}).

\bibitem[{\citenamefont{Affleck and Ludwig}(1991)}]{AffleckAndLudwig}
\bibinfo{author}{\bibfnamefont{I.}~\bibnamefont{Affleck}} \bibnamefont{and}
  \bibinfo{author}{\bibfnamefont{A.~W.~W.} \bibnamefont{Ludwig}},
  \bibinfo{journal}{Phys. Rev. Lett.} \textbf{\bibinfo{volume}{67}},
  \bibinfo{pages}{161} (\bibinfo{year}{1991}).

\bibitem[{\citenamefont{Laflorencie et~al.}(2006)\citenamefont{Laflorencie,
  Sorensen, Chang, and Affleck}}]{Ian1}
\bibinfo{author}{\bibfnamefont{N.}~\bibnamefont{Laflorencie}},
  \bibinfo{author}{\bibfnamefont{E.~S.} \bibnamefont{Sorensen}},
  \bibinfo{author}{\bibfnamefont{M.-S.} \bibnamefont{Chang}}, \bibnamefont{and}
  \bibinfo{author}{\bibfnamefont{I.}~\bibnamefont{Affleck}},
  \bibinfo{journal}{Phys. Rev. Lett.} \textbf{\bibinfo{volume}{96}},
  \bibinfo{pages}{100603} (\bibinfo{year}{2006}).

\bibitem[{\citenamefont{Jacobsen and Saleur}(2008)}]{XXX}
\bibinfo{author}{\bibfnamefont{J.~L.} \bibnamefont{Jacobsen}} \bibnamefont{and}
  \bibinfo{author}{\bibfnamefont{H.}~\bibnamefont{Saleur}},
  \bibinfo{journal}{Phys. Rev. Lett.} \textbf{\bibinfo{volume}{100}},
  \bibinfo{pages}{087205} (\bibinfo{year}{2008}).

\bibitem[{\citenamefont{White et~al.}(1994)\citenamefont{White, Noack, and
  Scalapino}}]{White1994}
\bibinfo{author}{\bibfnamefont{S.~R.}~\bibnamefont{White}},
  \bibinfo{author}{\bibfnamefont{R.~M.}~\bibnamefont{Noack}}, \bibnamefont{and}
  \bibinfo{author}{\bibfnamefont{D.~J.}~\bibnamefont{Scalapino}},
  \bibinfo{journal}{Phys. Rev. Lett.} \textbf{\bibinfo{volume}{73}},
  \bibinfo{pages}{886} (\bibinfo{year}{1994}).

\bibitem{Snote} {We note that other choices of bipartition compatible with DMRG sweeping give similar results for Fig.~3.}


  \bibitem[{\citenamefont{Wolf}(2006)}]{2dfermion}
  \bibinfo{author}{\bibfnamefont{M.~M.}~\bibnamefont{Wolf}},
  \bibinfo{journal}{Phys. Rev. Lett.},
  \textbf{\bibinfo{volume}{96}}, \bibinfo{pages}{010404}
  (\bibinfo{year}{2006}).

\bibitem[{\citenamefont{Cramer et~al.}(2006)\citenamefont{Cramer, Eisert, and
  Plenio}}]{2dboson}
\bibinfo{author}{\bibfnamefont{M.}~\bibnamefont{Cramer}},
  \bibinfo{author}{\bibfnamefont{J.}~\bibnamefont{Eisert}}, 
  \bibinfo{author}{\bibfnamefont{M.~B.}~\bibnamefont{Plenio}}, \bibnamefont{and}
  \bibinfo{author}{\bibfnamefont{J.}~\bibnamefont{Dreissig}},
  \bibinfo{journal}{Phys. Rev. A} \textbf{\bibinfo{volume}{73}},
  \bibinfo{pages}{012309} (\bibinfo{year}{2006}).

\bibitem[{\citenamefont{Kozma et~al.}(2005)\citenamefont{Kozma, Hastings, and
  Korniss}}]{network}
\bibinfo{author}{\bibfnamefont{B.}~\bibnamefont{Kozma}},
  \bibinfo{author}{\bibfnamefont{M.~B.}~\bibnamefont{Hastings}}, \bibnamefont{and}
  \bibinfo{author}{\bibfnamefont{G.}~\bibnamefont{Korniss}},
  \bibinfo{journal}{Phys. Rev. Lett.} \textbf{\bibinfo{volume}{95}},
  \bibinfo{pages}{018701} (\bibinfo{year}{2005}).

\bibitem[{\citenamefont{Sheng et~al.}(2009)\citenamefont{Sheng, Motrunich, and
  Fisher}}]{BoseMetal}
\bibinfo{author}{\bibfnamefont{D.~N.} \bibnamefont{Sheng}},
  \bibinfo{author}{\bibfnamefont{O.~I.} \bibnamefont{Motrunich}},
  \bibnamefont{and} \bibinfo{author}{\bibfnamefont{M.~P.~A.}
  \bibnamefont{Fisher}}, \bibinfo{journal}{Phys. Rev. B}
  \textbf{\bibinfo{volume}{79}}, \bibinfo{pages}{205112}
  (\bibinfo{year}{2009}).

\end{thebibliography}

\end{document}